\documentclass[aip,
jcp,
preprint,
longbibliography,
nofootinbib
]{revtex4-1}

\usepackage[dvipdfmx]
{graphicx}
\usepackage{amssymb}
\usepackage{amsmath}
\usepackage{braket}
\usepackage{listings}
\usepackage{cases}
\usepackage{comment}
\usepackage[colorlinks=true,
urlcolor=blue,
anchorcolor=blue,
citecolor=blue,
filecolor=blue,
linkcolor=blue,
menucolor=blue,
linktocpage=true,
pdfproducer=medialab,
pdfa=true
,dvipdfmx
]{hyperref}

\def\slashchar#1{\setbox0=\hbox{$#1$} 
\dimen0=\wd0 
\setbox1=\hbox{/} \dimen1=\wd1 
\ifdim\dimen0>\dimen1 
\rlap{\hbox to \dimen0{\hfil/\hfil}} 
#1 
\else 
\rlap{\hbox to \dimen1{\hfil$#1$\hfil}} 
/ 
\fi}

\begin{document}

\title{Tensor decomposition methods for correlated electron pairs}
\date{\today}
\affiliation{The Institute for Solid State Physics, The University of Tokyo, 5-1-5 Kashiwanoha, Kashiwa, Chiba, 277-8581, Japan}
\author{Airi Kawasaki}
\email{a\_kawasaki@issp.u-tokyo.ac.jp}
\author{Osamu Sugino}
\email{sugino@issp.u-tokyo.ac.jp}

\begin{abstract}
We analyze wave functions constructed as a sum of product of two-electron functions, or as a polynomial of geminals, to investigate their ability to represent the ground state of a strongly correlated few-body system. The known difficulty associated with variational determination of the total energy is overcome by applying a tensor decomposition method called Waring decomposition. Convergence speed of the total energy is compared for various polynomial types. The result provides information bridging between geminal product wave functions and the full-CI in the strongly correlated regime, thereby enriching knowledge on the hierarchy of molecular orbital theories of electron pairs.
\end{abstract}


\maketitle

\section{Introduction}

The concept of Fermion pair has long been used in various fields of quantum physics to characterize interacting Fermion systems. Typical examples can be seen in the molecular orbital theory of chemical bonds, interacting boson models in nuclear physics, and the BCS theory of superconductivity. In the field of quantum chemistry, already in 1950-1960s, chemical bonds were represented by assigning a two-electron function called geminal \cite{shull1959natural} to each electron pair and the many-body wave function was constructed using an antisymmetrized product of the geminals (APG) \cite{kutzelnigg1964w}. Because of the complicated form for the total energy, however, property of the APG wave functions has been mostly investigated by restricting the degrees of freedom, as can be seen either in the theory of antisymmetrized product of strongly orthogonal geminal (APSG) \cite{hurley1953molecular} to impose strongly orthogonal condition among the geminals or in the theory of antisymmetrized product of interacting geminals (APIG) \cite{silver1969natural} to enforce two electrons to occupy the same spatial orbital. Nevertheless, until recently, there has been continuous effort to relax the restrictions for better description of the correlation effect. The resulting hierarchy of the methods, ranging from APSG to an unrestricted APG, was discussed recently in the literatures \cite{tecmer2014assessing,limacher2016new}. Their ability to represent the ground state wave functions has been investigated for small molecules, but investigation for strongly correlated systems can be scarcely seen in the literatures although the study in that direction is also important for enriching our knowledge on the hierarchy. In our study, we will exclude the discussion on the antisymmetrized geminal powers (AGP) wave function augmented with a Jastrow type correlation factor, known as the JAGP theory \cite{2003JChPh.119.6500C}: JAGP is another paired electron theory for a Monte Carlo simulation of correlated electron systems developed rather distinctly from the molecular orbital theory of chemical bonds.

In this context, the purpose of this study is to examine the unrestricted APG wave function and its extensions constructed as a superposition of APGs or as a polynomial of geminal functions in the strong correlation regime. We will investigate how APG and the extensions are close to the full-CI and how efficiently they can be let approach the full-CI. By this, we try to reveal how the concept of electron pair is effective in representing the ground state wave function. To enable the study, we derive a formula to transform a polynomial of geminals to a linear combination of AGPs, or a AGP-CI wave function, since for the latter, the total energy scheme was derived and tested \cite{onishi1966generator,2015PhRvA..91f2504U,kawasaki2016four}. The transformation is based on a mathematical theory of tensor decomposition called Waring decomposition \cite{waring,oeding2013eigenvectors,comon1996decomposition,alexander1995polynomial}. In this study, we will pay attention to several known formulae for the Waring decomposition, which are used to transform several specific forms for the polynomial. General formulae for the Waring decomposition will be required to accomplish systematic study, but constructing a general formula is known to be difficult \cite{oeding2013eigenvectors}.

We will use a simple one-dimensional Hubbard model for our study, although application to molecules is rather straightforward; in fact, linear combination of AGP was tested for small molecules in the past \cite{2015PhRvA..91f2504U}.

\section{Formulation}
\label{Formulation}

\subsection{Geminal product wave functions and waring decomposition}

The two-electron function, called geminal, is defined as
\begin{eqnarray}
\hat{F}  \equiv \sum_{ab}^{M_{\text{base}}}F_{ab}c^{\dag}_{a}c^{\dag}_{b}, \label{geminal}
\end{eqnarray}
where $F_{ab}$ is an antisymmetric matrix and $M_{\text{base}}$ is the number of bases. Hereafter, we assume it to be a real matrix because we focus only on finite systems. By taking $n/2$-th power of the geminal as
\begin{eqnarray}
\ket{F} =  \hat{F} ^{\frac{n}{2}} \ket{0} \propto \left. \exp [\hat{F}t] \right| _{t^{\frac{n}{2}}}, \label{agp}
\end{eqnarray}
where $n$ is the number of electrons and $| _{t^m}$ means to take $m$-th order coefficient of a polynomial, antisymmetrized geminal power (AGP) is formed \cite{1965JMP.....6.1425C}. The right-hand side of Eq.\ (\ref{agp}) indicates that the AGP wave function is the BCS wave function of fixed number of electrons. Note that an antisymmetric matrix is transformed to a canonical form, or applied with real Schur decomposition, as
\begin{eqnarray}
F_{ab}=\sum_{k}V_{ak}V_{b\bar{k}}\epsilon_{k\bar{k}}
\end{eqnarray}
where $V$ is a real orthogonal matrix and $\epsilon$ is a real $2\times 2$ antisymmetric matrix of the form
\begin{eqnarray}
  \left(
    \begin{array}{cc}
      0 & a \\
      -a & 0 
    \end{array}
  \right) \label{occupation}
\end{eqnarray}
with square of $a$ corresponding to ``occupation number'' of the canonical state labeled by $k$ and its conjugate state labeled by $\bar{k}$. Then, the geminal is decomposed into the canonical ones as
\begin{eqnarray}
\hat{F}=\sum_{k}a^{\dag}_{k} a^{\dag}_{\bar{k}} \epsilon_{k\bar{k}} \equiv \sum_{k} \hat{F}_k
\end{eqnarray}
where
\begin{eqnarray}
a^{\dag}_{k}  \equiv \sum_{a} c^{\dag}_{a}V_{ak},
\end{eqnarray}
and then the AGP wave function (Eq.\ (\ref{agp})) is rewritten as
\begin{eqnarray}
\hat{F}^{n/2}\ket{0} = \sum_{1\leq i_1 < i_2 < \cdots < i_{n/2} \leq M_{\text{base}}} \hat{F}_{i_1}\hat{F}_{i_2}\cdots \hat{F}_{i_{n/2}} \ket{0}. \label{agp2}
\end{eqnarray}
Therefore, AGP wave function is comprised of $n/2$ electron pairs accommodated in the canonical states.

The AGP wave function was used to develop the generator coordinate method (GCM) in nuclear physics and was given an analytic form for the overlap and the Hamiltonian matrix element \cite{onishi1966generator}; for example, overlap between AGPs consisting of different geminals $F[\lambda]$ and $F[\mu]$ is written as
\begin{eqnarray}
\braket{F[{\lambda}]|F[{\mu}]} = \left.\exp\left(  \frac{1}{2}\mathrm{tr}\left[  \ln(1+F[{\mu}]F[{\lambda}]^{\dag}t^{2})\right] \right)\right| _{t^{\frac{n}{2}}} \label{norm} ,
\end{eqnarray}
where $\dag$ in the superscript means to take transpose.
This makes it possible to fully optimize the geminals to variationally determine the ground state wave function. The resulting wave function is an extension of the Hartree-Fock in that the electron pairs are treated independently with each other while intra pair repulsion is treated properly. To take the inter pair correlation into account, AGP wave functions can be linearly combined to form an AGP-CI wave function. It was shown that small number of AGPs was sufficient to achieve the full-CI limit for systems consisting of small number of electrons \cite{2015PhRvA..91f2504U,kawasaki2016four}. The small number of terms required for AGP-CI contrasts with much larger number of Slater determinants required for the full-CI. The difference in the number is originated from the fact that no orthonormality condition is applied to the former contrary to the latter; removal of the orthonormal restriction is thus important to compactly represent the wave function.

APG wave function is described as
\begin{eqnarray}
 \hat{F}[1]\hat{F}[2]\cdots\hat{F}[n/2]\ket{0} . \label{apg}
\end{eqnarray}
Contrary to AGP wave function (Eq.\ (\ref{agp})), compact analytic form is not known for the overlap and the Hamiltonian matrix element \cite{limacher2013new}. This fact has made it difficult to obtain the ground state wave function. In this context, we notice commutable nature of the geminal (Eq.\ (\ref{geminal})), which makes it possible to use the Waring decomposition formula for a monomial, known as the Fischer's formula \cite{fischer1994sums}, as
\begin{eqnarray}
&&\hat{F}[1]\hat{F}[2]\cdots\hat{F}[N] \nonumber \\ 
&&= \frac{1}{2^{N-1}N!}\sum_{I\subset\{ 1, 2, \cdots , N \}\setminus\{1\}}(-1)^{|I|}\left(\hat{F}[1]+\delta (I, 2)\hat{F}[2]+\cdots + \delta (I, N)\hat{F}[N]\right)^N, \label{fischer}
\end{eqnarray}
where $\delta (I, i)=-1$ when $i\in I$ and $=1$ when otherwise, to change the APG wave function (Eq.\ (\ref{apg})) into the sum of AGP wave functions. In Eq.\ (\ref{fischer}), $|I|$ means the number of elements contained in the set $I$.

\subsection{Linear combination of APG}

To consider the electron correlation effects missing in APG, we make a linear combination of APGs or APG-CI, which may also be called as a polynomial of geminals. The commutable nature of geminals also allows us to apply the Waring decomposition to transform the APG-CI wave function to a linear combination of AGPs like
\begin{eqnarray}
\sum_{i_1,i_2,\cdots ,i_M}^{i_1+\cdots +i_M =N}C_{i_1,i_2,\cdots ,i_M}\hat{F}[1]^{i_1}\hat{F}[2]^{i_2}\cdots \hat{F}[M]^{i_M} = \sum_{p=1}^{R}C_p\left(\sum_{k}a_{pk}\hat{F}[k]\right)^N \label{apg-ci} .
\end{eqnarray}
Note that, among those decompositions that satisfy Eq.\ (\ref{apg-ci}), those with the smallest number of terms appearing in the right-hand side is called Waring decomposition and the number $R$ is called Waring rank. In fact, it is not an easy task to find the Waring decomposition for a general polynomial; algorithms are under development\cite{oeding2013eigenvectors,carlini2012solution}. In this context, we use only those Waring decompositions whose formulae have been discovered and, with those decompositions, we try to relate the corresponding polynomial types with their ability to represent the wave function. We will hereafter call all that satisfy Eq.\ (\ref{apg-ci}) as a Waring decomposition if not ambiguous. Followings are the adopted Waring decompositions.

\subsubsection{Elementary symmetric polynomial} 

General form for the elementary symmetric polynomial is
\begin{eqnarray}
\sum_{1\leq i_1 < i_2 < \cdots < i_N \leq M}\hat{F}[i_1]\hat{F}[i_2]\cdots \hat{F}[i_N] 
&=& \left. \prod ^M _{i=1} \left( 1+ \hat{F}[i]t \right) \right|_{t^N} \nonumber \\ 
&\equiv & e_N(\hat{F}[1], \hat{F}[2], \cdots , \hat{F}[M]). \label{ele}
\end{eqnarray}
The Waring decomposition is known to be \cite{LEE201689}
\begin{eqnarray}
\frac{1}{2^{N}(M-N)N!}  \sum^{| I | \leq N/2}_{ I \subset \{ 1, 2, \cdots , M \} }  (-1)^{|I|}  \left(
\begin{array}{c} M-N/2 - |I| -1 \\
N/2 - |I|  
\end{array}\right) (M-2|I|) &  \nonumber \\
\times\left(\delta (I, 1)\hat{F}[1]+\delta (I, 2)\hat{F}[2]+\cdots + \delta (I, M)\hat{F}[M]\right)^N
\end{eqnarray}
for even $N$ and
\begin{eqnarray}
\frac{1}{2^{N-1}N!}     \sum^{| I | \leq (N-1)/2}_{ I \subset \{ 1, 2, \cdots , M \} }  (-1)^{|I|}   \left(
\begin{array}{c} M-(N-1)/2 - |I| -1 \\ \label{deco}
(N-1)/2 - |I|  
\end{array}\right)   & \nonumber  \\
\times\left(\delta (I, 1)\hat{F}[1]+\delta (I, 2)\hat{F}[2]+\cdots + \delta (I, M)\hat{F}[M]\right)^N 
\end{eqnarray}
for odd $N$. Explicitly for $N=3$ ($6$ electrons) and $M=4$ ($4$ sets of geminals), the decomposition is
\begin{eqnarray}
&&\hat{F}[1]\hat{F}[2]\hat{F}[3]+\hat{F}[1]\hat{F}[2]\hat{F}[4]+\hat{F}[1]\hat{F}[3]\hat{F}[4]+\hat{F}[2]\hat{F}[3]\hat{F}[4] \nonumber\\
=\frac{1}{24}&&\left[ 2( \hat{F}[1] + \hat{F}[2] + \hat{F}[3] + \hat{F}[4] )^3 -( -\hat{F}[1] + \hat{F}[2] + \hat{F}[3] + \hat{F}[4] )^3 \right. \nonumber \\
&&- ( \hat{F}[1] - \hat{F}[2] + \hat{F}[3] + \hat{F}[4] )^3 - ( \hat{F}[1] + \hat{F}[2] - \hat{F}[3] + \hat{F}[4] )^3  \nonumber\\
&&\left. -( \hat{F}[1] + \hat{F}[2] + \hat{F}[3] - \hat{F}[4] )^3 \right] \label{ee}
\end{eqnarray}
Physical picture behind taking $e_N$ is to distribute $N$ electron pairs into $M$ geminal functions (geminals) without allowing doubly occupying a geminal. All the geminals appear symmetrically and are therefore equivalently treated. Note the similar structure shared by Eq.\ (\ref{ele}) and Eq.\ (\ref{agp2}), namely $e_N(\hat{F}[1],\hat{F}[2],\cdots ,\hat{F}[N])$ and $e_N(\hat{F}_1,\hat{F}_2,\cdots ,\hat{F}_{M_{\text{base}}})$, respectively. They are, however, different in that $\hat{F}$'s appearing in the former are independent from each other while those in the latter are restricted to be orthogonal to each other; $\hat{F}_i$ is rank-1 matrix, i.e., $\hat{F}_i^2=0$. The former thus has larger degrees of freedom and can be regarded as an extension of the latter.

\subsubsection{Complete homogeneous symmetric polynomial} 

General form for the complete homogeneous symmetric polynomial is
\begin{eqnarray}
 \sum_{1\leq i_1 \leq i_2 \leq \cdots \leq i_N \leq M}\hat{F}[i_1]\hat{F}[i_2]\cdots \hat{F}[i_N] \equiv h_N(\hat{F}[1], \hat{F}[2], \cdots , \hat{F}[M]) .
\end{eqnarray}
As far as we know, there is no known form for the Waring decomposition, so that we use the Fischer's formula (Eq.\ (\ref{fischer})) term by term. Explicitly for $N=3$ and $M=4$, the decomposition is
\begin{eqnarray}
&&\hat{F}[1]^3 + \hat{F}[1]^2 \hat{F}[2] + \hat{F}[1]  \hat{F}[2]^2 + \hat{F}[2]^3 + \hat{F}[1]^2 \hat{F}[3] +\hat{F}[1] \hat{F}[2]\hat{F}[3] + \hat{F}[2]^2 \hat{F}[3] + \hat{F}[1]\hat{F}[3]^2 \nonumber \\
&&+ \hat{F}[2]\hat{F}[3]^2 + \hat{F}[3]^3 + \hat{F}[1]^2 \hat{F}[4] + \hat{F}[1] \hat{F}[2] \hat{F}[4] +  \hat{F}[2]^2 \hat{F}[4] + \hat{F}[1] \hat{F}[3] \hat{F}[4]+  \hat{F}[2] \hat{F}[3]\hat{F}[4] \nonumber \\
&&+ \hat{F}[3]^2 \hat{F}[4] +\hat{F}[1] \hat{F}[4]^2 +  \hat{F}[2] \hat{F}[4]^2 + \hat{F}[3] \hat{F}[4]^2 + \hat{F}[4]^3 \nonumber\\
&=&\frac{1}{6}\left[3 \hat{F}[1]^3 + 3 \hat{F}[2]^3 + 3 \hat{F}[3] ^3 + 3 \hat{F}[4]^3 + (\hat{F}[2] + \hat{F}[3]  + \hat{F}[4])^3\right. \nonumber \\
&&\left.+ (\hat{F}[1] +\hat{F}[3]  + \hat{F}[4])^3 + (\hat{F}[1] +\hat{F}[2]+ \hat{F}[4])^3 + (\hat{F}[1] +\hat{F}[2] +\hat{F}[3] )^3\right] . \label{eh}
\end{eqnarray}
By taking $h_N$ we symmetrically distribute $N$ electron pairs into $M$ geminals allowing a multiple occupation of a geminal. Note that the difference between $e_N$ and $h_N$ is reflected in the sign of the transformed AGPs, as can be seen in Eqs.\ (\ref{ee}, \ref{eh}).

\subsubsection{Permanent polynomial} 

Permanent polynomial is constituted by $m^2$ geminals, $F[{a,b}]$ with $1\leq a ,b \leq m $, as
\begin{eqnarray}
 \sum_{\sigma\in S_m}\hat{F}[{1,\sigma (1)}]\hat{F}[{2,\sigma (2)}]\cdots \hat{F}[{m,\sigma (m)}] \equiv \text{perm}_m(\hat{F}[1,1],\cdots , \hat{F}[m,m]), 
\end{eqnarray}
when $m$ is taken to be equal to the number of electron pairs $N$. The Waring decomposition is not known but can be made by applying the Fischer's formula to the Ryser-Glynn formula \cite{glynn2010permanent}
\begin{eqnarray}
\text{perm}_m(\hat{F}[1,1],\cdots , \hat{F}[m,m]) = \frac{1}{2^{m-1}}\sum_{\epsilon =\{ -1,1\} , \epsilon_1 =1}\prod_{1\leq i \leq m}\sum_{1\leq j \leq m}\epsilon_i \epsilon_j \hat{F}[{i,j}] \label{perm}
\end{eqnarray}
following Refs. \cite{teitler2014geometric,landsberg2017permanent}. For $N=3$, Eq.\ (\ref{perm}) becomes
\begin{eqnarray}
\frac{1}{4}&&\left[ (\hat{F}[1,1] + \hat{F}[1, 2] +\hat{F}[1, 3] )(\hat{F}[2,1] + \hat{F}[2, 2] +\hat{F}[2, 3] )(\hat{F}[3,1] + \hat{F}[3, 2] +\hat{F}[3, 3] ) \right. \nonumber\\
-&&(\hat{F}[1,1] + \hat{F}[1, 2] -\hat{F}[1, 3] )(\hat{F}[2,1] + \hat{F}[2, 2] -\hat{F}[2, 3] )(\hat{F}[3,1] + \hat{F}[3, 2] -\hat{F}[3, 3] )\nonumber\\
-&&(\hat{F}[1,1] - \hat{F}[1, 2] +\hat{F}[1, 3] )(\hat{F}[2,1] - \hat{F}[2, 2] +\hat{F}[2, 3] )(\hat{F}[3,1] - \hat{F}[3, 2] +\hat{F}[3, 3] ) \nonumber\\
+&&\left.(\hat{F}[1,1] - \hat{F}[1, 2] -\hat{F}[1, 3] )(\hat{F}[2,1] - \hat{F}[2, 2] -\hat{F}[2, 3] )(\hat{F}[3,1] - \hat{F}[3, 2] -\hat{F}[3, 3] ) \right] ,\nonumber\\
\end{eqnarray}
which can be recognized as the terms consisting of four APG's. Note that the degrees of freedom are smaller than those of linear combination of four independent APG's such as
\begin{eqnarray}
\sum_{i=1}^4 \hat{F}[1, i]\hat{F}[2, i]\hat{F}[3, i].
\end{eqnarray}
%

\subsubsection{Determinant polynomial} 

Likewise, we can introduce determinant polynomial as
\begin{eqnarray}
 \sum_{\sigma\in S_m}\text{sgn}(\sigma)\hat{F}[{1,\sigma (1)}]\hat{F}[{2,\sigma (2)}]\cdots \hat{F}[{m,\sigma (m)}] \equiv \text{det}_m(\hat{F}[1,1],\cdots , \hat{F}[m,m]) \label{det}
\end{eqnarray}
One can use formula for Derksen \cite{derksen2016nuclear} followed by applying the Fischer's formula to get the Waring decomposition. The Derksen formula for $m=3$ is given as
\begin{eqnarray}
 \frac{1}{2} \Bigl[&& (F[1,3]+F[1,2])(F[2,1]-F[2,2])(F[3,1]+F[3,2]) \nonumber\\
&&+ (F[1,1]+F[1,2])(F[2,2]-F[2,3])(F[3,2]+F[3,3]) \nonumber\\
&& + 2F[1,2](F[2,3]-F[2,1])(F[3,3]+F[3,1]) \nonumber\\
&&+ (F[1,3]-F[1,2])(F[2,2]+F[2,1])(F[3,2]-F[3,1]) \nonumber\\
&&  + (F[1,1]-F[1,2])(F[2,3]+F[2,2])(F[3,3]-F[3,2]) \Bigr].
\end{eqnarray}
%

\subsection{Total energy formula} 

The total energy formula was already shown in the forgoing researches \cite{onishi1966generator,2015PhRvA..91f2504U,kawasaki2016four}. Therefore, we show a summary here. The overlap was shown in Eq.\ (\ref{norm}). The matrix element of the one-body term can be obtained using the commutation relation derived by Onishi and Yoshida \cite{onishi1966generator},
\begin{eqnarray}
\left[  c_{\alpha},\exp\left( \hat{F}\right)  \right]  =\sum_{\delta}F_{\alpha\delta}c_{\delta}^{\dag}%
\exp\left( \hat{F}\right),
\label{OYc}%
\end{eqnarray}
as
\begin{eqnarray}
\left.\bra{tF[{\lambda}]}c^{\dag}_{a}c_{b}\ket{tF[\mu]} \right| _{t^{N}}= \left(\frac{F[{\mu}]F[{\lambda}]^{\dag}t^{2}}{1+F[{\mu}]F[{\lambda}]^{\dag}t^{2}}\right)_{ba} \left.\exp\left(  \frac{1}{2}\mathrm{tr}\left[  \ln(1+F[{\mu}]F[{\lambda}]^{\dag}t^{2})\right] \right)\right| _{t^{N}} .\label{one}
\end{eqnarray}
The two-body term is obtained as
\begin{eqnarray}
\left.\bra{tF[{\lambda}]}c^{\dag}_{p}c^{\dag}_{q}c_{s}c_{r}\ket{tF[{\mu}]} \right| _{t^{N}} &=& \left( \left[\frac{F[{\mu}]F[{\lambda}]^{\dag}t^{2}}{1+F[{\mu}]F[{\lambda}]^{\dag}t^{2}}\right]_{rp}\left[\frac{F[{\mu}]F[{\lambda}]^{\dag}t^{2}}{1+F[{\mu}]F[{\lambda}]^{\dag}t^{2}}\right]_{sq} \right.\nonumber \\
&& -\left[\frac{F[{\mu}]F[{\lambda}]^{\dag}t^{2}}{1+F[{\mu}]F[{\lambda}]^{\dag}t^{2}}\right]_{rq}\left[\frac{F[{\mu}]F[{\lambda}]^{\dag}t^{2}}{1+F[{\mu}]F[{\lambda}]^{\dag}t^{2}}\right]_{sp} \nonumber\\
&& \left. +\left[\frac{t}{1+F[{\mu}]F[{\lambda}]^{\dag}t^{2}}F[{\mu}]\right]_{rs}\left[F[{\lambda}]^{\dag}\frac{t}{1+F[{\mu}]F[{\lambda}]^{\dag}t^{2}}\right]_{qp} \right)\nonumber\\
&&\times\left.\exp\left(  \frac{1}{2}\mathrm{tr}\left[  \ln(1+F[{\mu}]F[{\lambda}]^{\dag}t^{2})\right] \right)\right| _{t^{N}} .\label{two}
\end{eqnarray}
To derive Eq.\ (\ref{one}), one can use Eq.\ (\ref{OYc}) to get
\begin{eqnarray}
c_{\alpha} \ket{tF} =  \sum_{\gamma}tF_{\alpha\gamma}c^{\dag}_{\gamma} \ket{tF},
\end{eqnarray}
and thus
\begin{eqnarray}
\bra{tF[{\lambda}]}c^{\dag}_{\alpha}c_{\beta}\ket{tF[{\mu}]} & = & \bra{tF[{\lambda}]}c^{\dag}_{\alpha}\sum_{\gamma}tF[{\mu}]_{\beta\gamma}c^{\dag}_{\gamma}\ket{tF[{\mu}]} \nonumber\\
& = & \sum_{\gamma}F[{\mu}]_{\beta\gamma}\frac{\partial}{\partial F[{\mu}]_{\alpha\gamma}}\braket{tF[{\lambda}]|tF[{\mu}]}.
\end{eqnarray}
Eq.\ (\ref{two}) can be derived in a similar way by relating the left-hand side with the derivative of the overlap; in that case it is easier to use the relation
\begin{eqnarray}
\frac{1}{t}\frac{\partial}{\partial F[{\mu}]_{cd}}\braket{tF[{\lambda}]|tF[{\mu}]} = \bra{tF[{\lambda}]}c^{\dag}_{c}c^{\dag}_{d}\ket{tF[{\mu}]}\label{bibun}
\end{eqnarray}
than to directly differentiating the formula by $F$.

\subsection{Hubbard model} 

Throughout this paper, we use a periodic one-dimensional Hubbard model with half-filling taking the value of $10$ for the Hubbard $U$ over the overlap $t$. The total energy is obtained by fully and independently optimizing all the geminals using the conjugate gradient (CG) method. No symmetry is assumed in the calculation. The optimization with CG tends to be slowed down by the non-convex nature of the total energy function rather than being trapped by local minima. Therefore, we have done the calculation using 100 initial guesses. The obtained total energy is compared with those obtained by exactly diagonalizing the Hamiltonian using the program package called H$\Phi$\cite{KAWAMURA2017180}. The resulting residual error in the total energy is compared.

\section{Result}
\label{Result}

\subsection{APG versus AGP-CI} 

\begin{figure}[htbp]
  \begin{center}
    \begin{tabular}{c}
    \hspace*{1.5cm} 
          \includegraphics[clip, width=10.0cm]{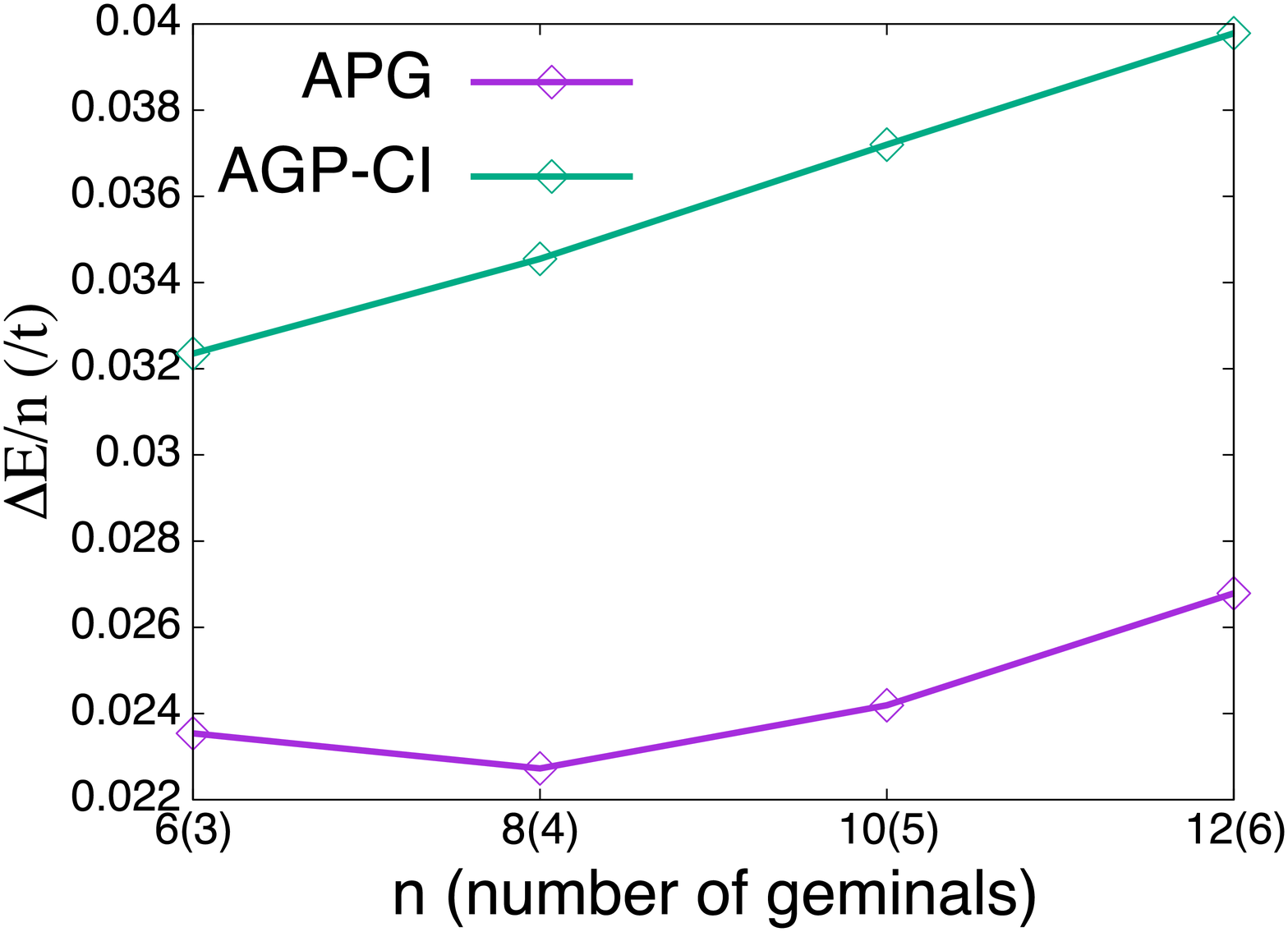}
          \hspace{1.6cm} 
     \end{tabular}
    \caption{(Color online) Residual error in the total energy per electron $\Delta E/n$ plotted against the number of electrons $n$.}
    \label{fig:zu1}
  \end{center}
\end{figure}

Now we investigate the property of APG and compare it with AGP-CI. For the sake of fair comparison, we use $M$ terms for AGP-CI so that the degrees of freedom are the same between the two calculations. FIG.\ \ref{fig:zu1} shows the residual error in the total energy. The error is always smaller for APG indicating its ability to more compactly represent the wave function. The error per electrons grows more moderately for APG than AGP-CI although the growth rate becomes comparable as $n$ is increased to 12. Note that the error of APG is not very small even for the six-electron system $n=6$. This indicates that the APG needs to be modified to very accurately describe the strongly correlated system specified by $U/t=10$. 

\subsection{Polynomial types versus accuracy} 

We investigate how the polynomial type affects the accuracy of the calculation. The calculation is done only for a system consisting of six electrons ($N=3$). The comparison of elementary symmetric polynomial ($e_3$), complete homogeneous symmetric polynomial ($h_3$), permanent ($\text{perm}_3$) and determinant ($\text{det}_3$) polynomial is shown in TABLE \ref{tab:tab1} and TABLE \ref{tab:tab2}. When using three geminals ($M=3$) for $e_3$ and $h_3$, the residual error in the total energy is $0.141t$ and $0.172t$, respectively. Note that $e_3$ consists of single term in this particular case and thus is the same as APG. The error is larger for $h_3$, which will be due to the effect of the multiply occupied geminals. Indeed, when the degrees of freedom are increased by introducing coefficients $C$ for each term in $h_3$ as
\begin{eqnarray}
\sum_{1\leq i_1 \leq i_2 \leq \cdots \leq i_N \leq M}C_{\{i_1, i_2, \cdots , i_N\}} \hat{F}[i_1]\hat{F}[i_2]\cdots \hat{F}[i_N]  \label{coe}
\end{eqnarray}
those coefficients with multiple occupancy are found small. The residual error is thereby reduced owing to the much larger degrees of freedom, but the accuracy does not appreciably exceed that of $e_3$. This fact also indicates minor contribution of the multiply occupied configurations like $\hat{F}[1]^3$.

When using nine geminals ($M=9$), the residual error is significantly reduced and $e_3$, $h_3$, and $\text{perm}_3$ yield approximately the same value of $\sim 10^{-5}t$. Contrary to the case of $M=3$, here the residual error is comparable for $e_3$ and $h_3$, which can be understood from the fact that the degrees of freedom are already large enough and the type of the polynomial is less important than for $M=3$. Yet, $\text{det}_3$ shows a superior value of the total energy. This indicates that Eq.\ (\ref{det}) is more suitable for the description of the total energy, but the reason is unclear though we have tried to figure it out.

\begin{figure}[htbp]
  \begin{center}
    \begin{tabular}{c}
    \hspace*{1.5cm} 
          \includegraphics[clip, width=10.0cm]{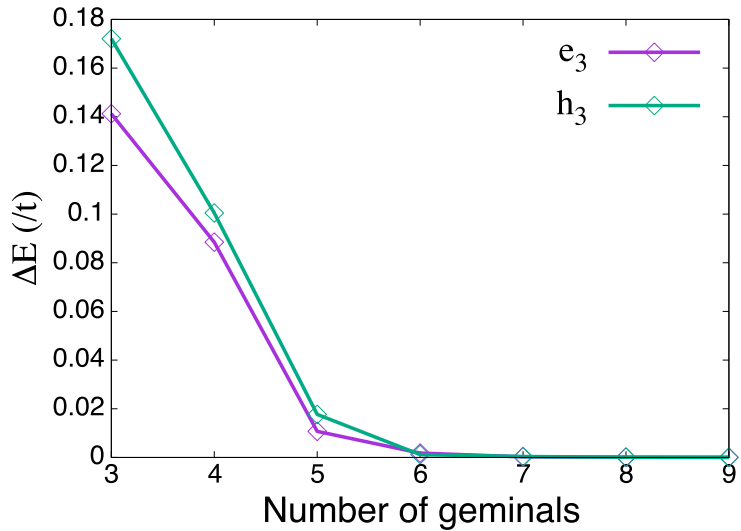}
          \hspace{1.6cm} 
     \end{tabular}
    \caption{(Color online) Residual error in the total energy $\Delta E$ plotted against the number of geminals $M$.}
    \label{fig:zu2}
  \end{center}
\end{figure}

We further compare in FIG.\ \ref{fig:zu2} how the residual error is reduced as the number of geminals is increased from $3$ to $9$ for $e_3$ and $h_3$. It also indicates that the error of $e_3$ is slightly smaller than that of $h_3$.

\begin{table}[htb]
  \begin{tabular}{|c||c |c | }\hline
     Polynomial type ($M=3$)& $ \Delta E/n$ $(/t)$ & Number of AGPs \\ \hline \hline
    $e_3$  & 0.141 &  4\\ \hline
    $h_3$  & 0.172 & 7 \\ \hline
    $h_3$ with $C$ (Eq.\ (\ref{coe}))& 0.140 &  19\\  \hline
  \end{tabular}
 \caption{Residual error in the total energy per electron $\Delta E/n$ and the number of AGPs generated after Waring decomposition.} 
 \label{tab:tab1}
\end{table}

\begin{table}[htb]
  \begin{tabular}{|c||c |c | }\hline
     Polynomial type ($M=9$) & $ \Delta E/n$ $(/t)$ & Number of AGPs \\ \hline \hline
    $e_3$   & 3.40E-05  & 10 \\ \hline
     $h_3$  &  4.35E-05  & 19 \\ \hline
     perm$_3$  & 4.71E-05 & 16 \\ \hline
     det$_3$  & 1.32E-07 &  20 \\ \hline
  \end{tabular}
 \caption{Residual error in the total energy per electron $\Delta E/n$ and the number of AGPs generated after Waring decomposition.} 
 \label{tab:tab2}
\end{table}

\section{SUMMARY AND CONCLUSION}
\label{conclusions}

We have investigated the APG wave function and its extensions, namely polynomial of geminals, to enrich the knowledge of the hierarchy on geminal theories in the strong correlation regime. This was done using one-dimensional Hubbard model ($U/t=10$) under the half-filling condition. First, we have found that the residual total energy error of APG has a moderate dependence on the number of electrons up to the 12 electrons system that we have investigated although the error is not so small, more than $0.02t$ per electron. By comparing the result with that obtained using the AGP-CI containing comparable degrees of freedom, we found that APG more efficiently represents the ground state.
We also found that the residual error of APG can be made very small for the 6 electrons system when using the polynomial containing 9 terms, indicating the ability to compactly represent the wave function. The error is found to depend on the polynomial type. Elementary symmetric polynomial is superior to complete homogeneous symmetric polynomial due to the lack of the multiply occupied configurations. Permanent and determinant polynomials, which also lack the multiple occupancy, are also examined. The former is as accurate as elementary symmetric polynomial while the latter is more accurate. The reason for the superior behavior of the determinant polynomial is unclear as far as we have investigated. Nevertheless, we expect that further efforts to relate the polynomial type and the accuracy will lead to a comprehensive explanation of the electron correlation in terms of the electron pair. The efforts will also help further sophisticate the molecular orbital theory of chemical bonds. We consider the present study is a step toward the goal. The computational cost is currently large and, in addition, grows exponentially with the number of electrons. It is worth trying to further reduce the cost by, for example, investigating whether the geminal can be localized in real space.

\begin{acknowledgements}
The authors thank the Supercomputer Center, the Institute for Solid State Physics, the University of Tokyo for the use of the facilities. Exact diagonalization was done by using H$\Phi$ \cite{KAWAMURA2017180}.
\end{acknowledgements}

\hspace{2cm}

\bibliography{cite}

\begin{thebibliography}{10}
\bibitem{waring}
Waring decomposition is also given other names in diverse areas of science,
  e.g., canonical decomposition, rank-1 decomposition, sum of powers
  decomposition.
\end{thebibliography}



\end{document}